\documentstyle[aps,preprint]{revtex}
\begin{document}
\draft
%\addtolength{\textheight}{5cm}
%\textheight 15in
%\textwidth 9in
%\baselineskip 8mm 
%\par 
\title{
Conductance renormalization and 
conductivity of a multi-subband Tomonaga-Luttinger model
}
\author{
Takashi Kimura
}
\address{
NTT Basic Research Laboratories, NTT Corporation, 
3-1 Morinosato-Wakamiya, 
Atsugi, Kanagawa 243-0198, Japan
}
\date{\today}
\maketitle
\begin{abstract}
We studied the conductance renormalization
and conductivity of multi-subband 
Tomonaga-Luttinger models with inter-subband interactions. 
We found that, as in single-band systems, the conductance of 
a multi-subband system with an arbitrary number of subbands 
is not renormalized due to interaction between electrons. 
We derived a formula for the conductivity in multi-subband models.
We applied it to a simplified case 
and found that inter-subband interaction enhances the conductivity, 
which is contrary to the intra-subband repulsive interaction, and that 
the conductivity is further enhanced for a larger number of subbands.
\end{abstract}
Recent studies of low-dimensional systems have brought to 
light many important properties. 
For instance, one-dimensional (1D) electron systems,   
in a low-energy regime, are described 
not by the Fermi liquid but by a Tomonaga-Luttinger (TL) liquid 
\cite{Tomonaga,Luttinger,TL}. 
Tomonaga-Luttinger liquids that 
include the effects of the multiple degrees of freedom,  
such as multi-chain TL models with the interchain hopping, 
have been extensively studied. 
In a bulk system, the interchain hopping 
between 1D TL chains is relevant, resulting in 
a strong-coupling regime that includes  
a spin gap and/or an enhanced superconducting correlation 
\cite{Fabrizio1993,Balents1996,Takashi1996,Lin1997}. 
The crossover from TL to Fermi Liquid has also been studied 
by including the interchain hopping \cite{Arrigoni1999}.
Regarding the transport properties, for example, 
a perfect transmission has been suggested in 
a two-chain system, 
reflecting the spin gap \cite{Takashi1995,Arrigoni1997}. 
The interchain conductivity\cite{Georges2000} 
and the Hall effect\cite{Lopatin2001} 
of a multi-chain system with the interchain hopping 
have also been discussed. 

TL liquids have been also studied in mesoscopic quantum wires, 
especially with respect to the transport properties.
The 1D Coulomb drag 
\cite{Komnik1998,Flensberg1998,Nazarov1998,Durganandini1999}
has been studied on 1D two-chain models coupled 
in a finite region \cite{Nazarov1998} 
or at a finite point(s) \cite{Komnik1998,Flensberg1998,Durganandini1999}. 
In these models, the interchain backward 
scattering process between electrons, 
which results in a strong-coupling regime, 
is essential for the occurrence of a perfect drag \cite{Nazarov1998}, 
a zero-bias anomaly \cite{Komnik1998}, or a power-law temperature 
dependence of the transconductance \cite{Flensberg1998}. 

Another TL system with multiple degrees of freedom is 
{\it a multi-subband TL model with inter-subband forward scattering}, 
where the inter-subband single-particle hopping is forbidden. 
Although this model 
is relevant to wide quantum wires with multi-subbands,
it has not been well studied for the transport properties, such as
conductance and conductivity. 
In a quantum wire, 
the long-range Coulomb interaction is not sufficiently screened  
and the forward scattering processes between electrons 
with a small momentum transfer play an important role, while 
the scattering processes with large momentum transfers of the order 
of the Fermi wave number(s), 
such as the backward, Umklapp, or inter-subband pair tunneling  
process, may be neglected. 
The ground state of 
the above multi-subband model is in a weak coupling regime
without the gapful excitation  
and is essentially different from the multi-chain model 
with the interchain hopping or the backward scattering, 
where the ground state is in a strong coupling regime. 

For single-band TL models, both the conductance of clean systems
\cite{Apel,Kane1992,Furusaki1993,TKA1996,Maslov1995,Ponomarenko1995,Kawabata1996,Shimizu1996}  
and the conductivity of dirty systems 
\cite{Luther,Fukuyama1993,Ogata1994,TKAE1994} 
have been studied. The models in 
refs.\cite{Maslov1995,Ponomarenko1995,Kawabata1996,Shimizu1996} 
include the effects of leading wires and show the absence of 
the conductance renormalization due to the electron-electron 
interaction, which is consistent with experiments \cite{Tarucha1995}. 
However, it is not so obvious 
whether the conductance renormalization of the multi-subband 
model is absent or not. 
For example, Liang and co-workers \cite{Liang2000} have 
experimentally found that, in a clean quantum wire, 
the conductance is smaller than the quantized conductance 
only in a high in-plane magnetic field, 
where the two inequivallent spin subbands cross the Fermi level. 
Hence, the conductance renormalization 
of a clean multi-subband TL model is also of interest. 

On the other hand, for a dirty single TL liquid, 
which can be realized in a long quantum wire  
where the wire length is longer than the 
mean-free path, a power-law temperature dependence of the conductivity 
was observed in experiment \cite{Tarucha1995}, which is consistent with the 
existing theory \cite{Luther,Fukuyama1993,Ogata1994}. 
If we consider a multi-subband system, in a two-subband system, 
then, as the author and co-workers \cite{TKAE1994} 
theoretically found, the inter-subband interaction 
enhances the conductivity 
even if the interaction is repulsive, 
contrary to the intra-subband repulsive interaction. 
In order to further clarify the multi-subband effect, 
the conductivity of a TL model 
with larger number of subbands should be investigated. 

%----------------------------------------------------
In this paper, we study the transport properties 
of the multi-subband TL model with 
the inter-subband forward scattering,   
neglecting the large momentum transfer processes, 
such as backward scatterings. 
We found that, as in single-band systems, 
the conductance of a clean multi-subband TL model 
with an arbitrary number of subbands
is not renormalized due to the interaction between electrons.         
We derived a formula based on the Mori formalism \cite{Mori,Gotze} 
for the conductivity of dirty multi-subband TL models. 
Applying the formula to a multi-subband model,  
we found that the inter-subband interaction 
enhances the conductivity for an arbitrary number of subbands, 
and that the conductivity is more 
enhanced for a larger number of subbands. 

{\it Conductance of a clean TL model.---} 
Let us start from a $N$-subband spinless TL model, 
which includes a spinful model  
as a special case; i.e., a spinless $2N$-subband model 
is equivalent to a spinful $N$-subband model. 
The spinless $N$-subband TL model 
can be represented as
\begin{eqnarray}
H&=&\sum_i^{N}\frac{1}{4\pi}\int\ dx \Big\{
v^i_N[\nabla \Theta^{i}_+(x)]^2
+v^i_J[\nabla \Theta^{i}_-(x)]^2
\Big\}\nonumber\\
&+&\sum_{i\neq j}^N
\frac{1}{4\pi}\int\ dx \Big\{
\frac{g^{ij}_N}{2}[\nabla \Theta^{i}_+(x)][\nabla \Theta^{j}_+(x)]
+\frac{g^{ij}_J}{2}[\nabla \Theta^{i}_-(x)][\nabla \Theta^{j}_-(x)]
\Big\}. \label{Hamiltonian}
\end{eqnarray}
Here $i$ or $j$ show the subband index. 
$\Theta^i_+$ is the phase variable for the $i$-th subband
 and $\Theta^i_-$  is its dual variable. 
$v^i_N \equiv v^i_F+g^i_4+g^i_2\equiv v^i/K^i$ 
and $v^i_J \equiv v^i_F+g^i_4-g^i_2\equiv v^i K^i$,  
where  $g^i_{2(4)}$ is the interaction 
parameter between electrons with the opposite 
(same) velocity direction in the $i$-th subband. 
$v^i_F$ is the Fermi velocity of the $i$-th subband 
and $v^i$ ($K^i$) is the velocity of the excitation 
(critical exponent) of the $i$-th subband. 
The inter-subband forward scatterings are included through 
$g^{ij}_N\equiv g^{ij}_4+g^{ij}_2$ 
and $g^{ij}_J\equiv g^{ij}_4-g^{ij}_2$,  
where $g^{ij}_{2}$ ($g^{ij}_{4}$)
is the interaction parameter between 
electrons with the opposite (same) velocity direction 
in the $i$-th and $j$-th subbands. 
The unit $e^2=\hbar=k_B=1$ is assumed throughout this paper. 
 
The conductance in the ballistic regime can be calculated 
by extending ref. \cite{Shimizu1996} for the single-band system. 
Following the usual manner \cite{Haldane1981}, 
the local current operator of the 
$i$-th subband is determined from the continuity equation for local density
$\rho^i(x)=\nabla \Theta^i_+(x)/(\sqrt{2}\pi)$ as 
\begin{eqnarray}
\frac{\partial{\hat j}^i(x) }{\partial x}=-\frac{\partial\rho^i(x)}{\partial t}
=-\frac{1}{\sqrt{2}\pi}\frac{\partial\Theta^i_+(x)}{\partial t}.
\end{eqnarray}
The dc mean current operator ${\hat j}_N^i$ is then given by
\begin{eqnarray}
{\hat j}_M^i&\equiv&\frac{1}{L}\int_0^L\ dx {\hat j}^i(x)\nonumber\\
&=&-\frac{i}{\sqrt{2}\pi L}\int_0^L\ dx \int_0^x \ dx^\prime 
[H,\nabla_x^\prime \Theta^i(x^\prime)]\nonumber\\
&=& \frac{1}{L}\Big(v^i_J {\hat J}_i
+\sum_{j(\neq i)}\frac{g^{ij}_J}{2}{\hat J_j}\Big),
\end{eqnarray}
where $L$ is the system length,
and ${\hat J}_i={\hat N}^i_1-{\hat N}^i_2$ is 
the operator for the difference 
between total number of particles 
of right-going electrons (${\hat N}^i_1$) and 
left-going ones (${\hat N}^i_2$) \cite{Haldane1981}.
On the other hand, the Hamiltonian can be rewritten as
\begin{eqnarray}
H &=& \sum_{k(\neq 0),i}\omega^i_k 
{\hat b}^{i\dagger}_k {\hat b}^i_k + \frac{\pi}{2L}
\sum_i^{N}\big[v^i_N 
{{\hat N}_i}^2 +v^i_J {\hat J}_i^2\big]\nonumber\\
&&+\frac{\pi}{4L}\sum_{i\neq j}^{N}
\big[g^{ij}_N {\hat N}_i {\hat N}_J +g^{ij}_J {\hat J}_i {\hat J_j}\big],
\end{eqnarray}
where ${\hat N}_i \equiv {\hat N}^i_1 +{\hat N}^i_2$ and 
$b^i_k$ is the annihilation operator of the boson 
with eigenenergy $\omega^i_k$ with some diagonalized index $i=1,..,N$.
Let $n^i_k$, $N_i$, and $J_i$ be the eigenvalues of 
${\hat b}^{i\dagger}_k {\hat b}^i_k$, $N_i$, and $J_i$, respectively.
The energy eigenvalue is given as
\begin{eqnarray}
E = \sum_{k(\neq 0),i} \omega^i_k n^i_k 
+ \frac{\pi}{2L}\sum_i^{N}\big[g^{ij}_N 
{N_i}^2 +g^{ij}_J J_i^2\big]
+\frac{\pi}{4L}\sum_{i\neq j}^{N}\big[v^i_N N_i N_j +v^i_J J_i J_j\big]
\end{eqnarray}
in the low-energy regime. 
Thus, the chemical potential difference between the right-going 
going electrons and the left-going electrons are 
obtained as 
\begin{eqnarray}
\delta\mu\equiv\mu_1-\mu_2=\frac{\partial E}{\partial N^i_{1}}
-\frac{\partial E}{\partial N^i_{2}}=
\frac{2\pi}{L}v^i_J J_i+
\frac{\pi}{L}\sum_{j(\neq i)}g^{ij}_J J_j {\rm\ (for\ all}\ i{\rm )}, 
\end{eqnarray}
where the subband index $i$ for $\delta\mu=\mu_1-\mu_2$ is 
omitted because both $\mu_1$ and $\mu_2$
must be the same for all subbands. 
$\delta\mu$ should equal the 
experimentally-observed chemical potential difference because 
both ends of the 1D system are connected to reservoirs  
(See \cite{Shimizu1996} for details). 
The conductance is readily obtained by 
using $j_M^i\equiv
(2v^i_J J_i
+\sum_{j(\neq i)}g^{ij}_J J_j)/2L$
(the eigenvalue of ${\hat j}_M^i$) as
\begin{eqnarray}
G=\frac{\sum_i j_M^i}{\delta\mu}=\frac{1}{2\pi}\times N.
\end{eqnarray}
Hence, the renormalization of the current and the chemical potential 
difference is completely canceled out like in a single-band system
\cite{Kawabata1996,Shimizu1996}, 
and hence the conductance of multi-subband systems with 
an arbitrary number of subbands is not renormalized 
due to the interaction between electrons 
and equals the quantized conductance (note that $\hbar=e^2=1$). 
From the present result, it is found that
the above-mentioned experiment\cite{Liang2000} in a magnetic field 
cannot be explained only by clean TL models, and the remaining possibilities 
\cite{zero,Ponomarenko1999} should be investigated. 

{\it Formula for the conductivity of a dirty TL model.---}
Here, we calculate the conductivity following 
G{\" o}tze and W{\" o}lfle\cite{Gotze} for the Mori formalism \cite{Mori}. 
We can calculate the subband-dependent relaxation time
in the second order of the impurity scattering.  
As a result, we obtain the conductivity $\sigma(T)$ as
\begin{eqnarray}
\sigma(T)&=&\sum_i\sigma_i(T)\nonumber\\
\sigma_i(T)&=&\sigma_{i0}F(\omega_F)/F(T),\nonumber\\
F(T) &\equiv&\frac{1}{T}\int_{-\infty}^{\infty}\ dt 
\langle\rho^i_{2k^i_F}(x=0,t)\rho^i_{2k^i_F}(x=0,t=0)\rangle, 
\label{F(T)}\\
\rho^i_{2k^i_F}(x)&\equiv& \Psi^\dagger_{i1}(x)\Psi_{i2}(x)+{\rm h.c},
\nonumber
\end{eqnarray}
Here, $\sigma_i(T)$ is the conductivity of $i$-th subband, 
$\sigma_{i0}=\sigma_i(T=\omega_F)=
n_i\tau_{i0}/m^*$ 
that of the free electrons, 
$\tau_{i0}=v^i_F/(n_i |u(2k^i_F)|^2)$, 
$\omega_F$ a high-frequency cutoff, $m^*$ the effective mass, 
$k^i_F$ $(n_i)$ the Fermi wave vector (the density of electrons) 
of the $i$-th subband, and $u(k)$ the impurity potential in momentum space.
$\Psi_{i1(2)}$ is the annihilation 
operator of right (left)-going electrons in the $i$-th 
subband. 
The Hamiltonian of Eq. \ref{Hamiltonian}
is written as
\begin{eqnarray}
H&=&\frac{1}{4\pi}\int\ dx \sum_{ij}^{N}\Big\{
H^+_{ij}[\nabla \Theta^{i}_+(x)]
[\nabla \Theta^{j}_+(x)]+
H^-_{ij}[\nabla \Theta^{i}_-(x)]
[\nabla  \Theta^{j}_-(x)]\Big\}, 
\end{eqnarray}
where $H^{+(-)}_{ii}=v^i_{N(J)}$ 
and $H^{+(-)}_{ij}=g^{ij}_{N(J)}/2$ for ($i\neq j$).
It is not so straightforward to find 
a linear transformation \cite{Nagaosa1993}, which diagonalizes 
both $H^{+}_{ij}$ 
and $H^{-}_{ij}$, although
we can diagonalize the Hamiltonian in principle, 
keeping the commutation relation 
$[\Theta^i_+(x),\frac{d \Theta^j_-(y)}{dy}]=-2\pi\delta_{ij}\delta(x-y)$.
However, if 
$g^{ij}_2=g^{ij}_4$ (i.e., $g^{ij}_J=0$), 
$H^-_{ij}$ is already diagonal and 
can be transformed to a matrix 
$H^{-\prime}_{ij}=v^1_J\delta_{ij}$ 
by a transformation 
$\Theta^i_+=\sqrt{v^i_J/v^1_J}\Theta^{i\prime}_+$, 
whereas the $\Theta_+$ part is simultaneously transformed as
$\Theta^i_-=\sqrt{v^1_J/v^i_J}\Theta^{i\prime}_-$,
$H^{+\prime}_{ij}= H^+_{ij}\sqrt{v^i_J v^j_J}/v^1_J$. 
By using a unitary matrix 
$U_{ij}\equiv({\vec u_1}, {\vec u_2},...,{\vec u_N})$, where  
${\vec u_i}$ is the $i$-th eigenvector of $H^{+\prime}_{ij}$ 
with the eigenvalue ${\tilde v}^i_N$, 
we can diagonalize $H^{+\prime}_{ij}$ 
as ${\tilde H}^{+}_{ij}\equiv\sum_{km} U^{-1}_{ik}H^{+\prime}_{km}U_{mj}={\tilde v}^i_N\delta_{ij}$ by a unitary transformation
$\Theta^{i\prime}_+=\sum_j U_{ij}{\tilde \Theta}^j_+$,  
whereas $H^{-\prime}_{ij}$ remains unchanged by the transformation
because $H^{-\prime}_{ij}$ is proportional to the unit matrix. Here, 
we should note that the condition $g^{ij}_2=g^{ij}_4$ 
is physically natural, because it holds whenever we
assume an effective Hamiltonian where only the total charge density is
coupled \cite{Haldane1981}. 
Since the Hamiltonian is now diagonalized, 
the density-density correlation functions can be calculated as
\begin{eqnarray}
\langle\rho^i_{2k^i_F}(0,t)\rho^i_{2k^i_F}(0,0)\rangle 
&\propto& {\rm exp}\Big[-2\sum_j {\tilde K}_j U_{ij}^2 
\int_0^{\infty}\frac{d\omega}{\omega}e^{-\omega/\omega_F}
\nonumber\\ &&
\times\Big\{
{\rm tanh}\Big(\frac{\omega}{2T}\Big)\Big(1-{\rm cos}(\omega T)\Big)
+i {\rm sin}(\omega T)\Big\}\Big]\nonumber\\
&\propto& \prod_j \Big[\frac{1+i\omega_F t}{\pi T t} 
{\rm sinh}\Big(\frac{\pi}{Tt}\Big)\Big]^{-2{\tilde K}_j U_{ij}^2},
\end{eqnarray}
where ${\tilde K}_i=\sqrt{v^1_J/{\tilde v}^i_J}$ 
is the critical exponent of 
${\tilde \Theta}^{i\prime}_+$. 
Finally, the formula for the conductivity of $i$-th subband 
is obtained by performing the time integral in Eq. \ref{F(T)} as
\begin{eqnarray}
\sigma_i(T)=\sigma_{i0}
\Big(\frac{T}{\omega_F}\Big)^{2\big(1-\sum_j{\tilde K}_j U_{ij}^2\big)}.
\end{eqnarray}

{\it Conductivity of an $N$-subband TL model.---}
We apply the above formula for a simplified 
case with $N$ spin-full electron subbands, where 
the calculation can be analytically performed for arbitrary $N$. 
We assume the intra- or inter-subband spin-independent 
interactions and the Fermi velocities 
are independent of the subband 
($g^i_2=g^i_4\equiv \pi v_F g/2$,  
$g_2^{ij}=g_4^{ij} \equiv \pi v_F g^\prime /2$,  and
$v_{iF}^\uparrow=v_{iF}^\downarrow\equiv v_F$ for all $i,j$),
whereas the Fermi wave number $k^i_F$ and the density of electrons $n_i$ 
naturally depend on $i$. 
The Hamiltonian can be written as
\begin{eqnarray}
H&=&\frac{v_F}{4\pi}\int\ dx 
\sum_i^N\Big\{(1+g)[\nabla \theta^i_+(x)]^2
+[\nabla  \theta^i_-(x)]^2 \Big\}\nonumber\\
&+&\frac{v_F g^\prime}{4\pi}\int\ dx 
\sum_{i\neq j}^N[\nabla \theta^i_+(x)][\nabla \theta^j_+(x)]\\
&+&\frac{v_F}{4\pi}\int\ dx 
\sum_i^N \Big\{ [\nabla \phi^{i}_+(x)]^2
+[\nabla \phi^{i}_-(x)]^2 \Big\}. \nonumber
\end{eqnarray}
Here $\theta^i_+$ ($\phi_+^i$) is the phase variable for the charge 
(spin) degree of freedom of the $i$-th subband, and $\theta_-^i$ ($\phi_-^i$) 
is its dual variable. 
The Hamiltonian matrix $H^+_{ij}$ 
($H^+_{ii}=1+g$, $H^+_{ij}=g^\prime$ for $i\neq j$)
has eigenvectors, such that  
$u_1=(1,1,...,1)/\sqrt{N}$, $u_2=(1,-1,0,...,0)/\sqrt{2}$, 
..., $u_i=(1,1,...,1-i(i$-th$),0,...,0)/\sqrt{i(i-1)}$, ...,
$u_N=(1,1,....,1,1-N)/\sqrt{N(N-1)}$, 
where $u_1$ has the eigenvalue $1+g+(N-1)g^\prime$ 
and the other eigenvectors
$u_{2\sim N}$ have the same eigenvalue $1+g-g^\prime$. 
One can perform a unitary transformation 
by using the eigenvectors to diagonalize $H^+_{ij}$
and finally obtain the conductivity as
\begin{eqnarray}
\sigma_i(T)&=&\sigma_{i0}
\Big(\frac{T}{\omega_F}\Big)^{2(1-K)},\nonumber\\
K&\equiv&\frac{1}{N}\frac{1}{\sqrt{1+g+(N-1)g^\prime}}
+\Big(1-\frac{1}{N}\Big)
\frac{1}{\sqrt{1+g-g^\prime}}.\label{K1}
\end{eqnarray}
For $N=2$, one can reproduce the result of ref. \cite{TKAE1994}.
Some interesting properties of the conductivity
can be found in Eq. \ref{K1}.
First, as in the single-band case, 
$\frac{\partial K}{\partial g}<0$ always holds, 
where the repulsive interaction enhances 
the $2k_F$ charge density wave (CDW) correlation, resulting in 
the reduction of the conductivity.
More interestingly, $\frac{\partial K}{\partial |g^\prime|}>0$
always holds for arbitrary subband numbers. 
This means that the inter-subband interaction, being independent of its sign,  
enhances the conductivity.
This is not so trivial but 
may be understood by the discordance between 
the wave numbers of the CDW correlations of different subbands. 
Namely, the inter-subband interaction 
disturbs the CDW correlation of each subband 
because of the discordance, 
and the effect of the disturbance should naturally 
be independent of the sign of the inter-subband interaction. 
On the other hand, $\frac{\partial K}{\partial N}>0$ also always holds, and  
thus the conductivity is monotonically enhanced 
as a function of the number of subbands. This is because
the inter-subband interaction, which enhances the conductivity, works 
more significantly for a larger number of subband. 
If we consider the large-$N$ (2D-like) behavior, which can be 
examined only when $g^\prime >0$,  
the critical exponent $K$ tends to $(1-1/N)/\sqrt{1+g-g'}+O(N^{-3/2})$ 
and the resulting conductivity is the same as in the single band system 
with a renormalized intra-subband interaction $g-g^\prime$.  
If one compares our result with that of 
ref. \cite{Brandes} based on the Fermi liquid ($\sigma(T)\propto T^{1/N}$), 
there is a qualitative consistency in the sense that the conductivity 
is an increasing function of $N$. 

In conclusion, we found that 
the conductance of clean systems 
with an arbitrary number of subbands 
is not renormalized due to the interaction between electrons.
We also found that the conductivity of a dirty multi-subband model 
is enhanced by the inter-subband interaction 
(contrary to the intra-subband repulsive interaction) 
independent of its sign and the number of subbands 
and that it is more enhanced for a larger number of subbands. 
The present results may be observed in future experiments on wide and long 
quantum wires that have multi-1D subbands.

The author deeply acknowledges 
Prof. Kazuhiko Kuroki for useful suggestions 
and Prof. Hideo Aoki for valuable discussions. 
He also thanks Dr. Hideaki Takayanagi for his encouragement. 

\end{document}